\documentstyle[tighten,floats,12pt,prd,aps,eqsecnum]{revtex}

\newcommand{\beq}{\begin{equation}}
\newcommand{\eeq}{\end{equation}}

\def\lap{\lower.5ex\hbox{$\; \buildrel < \over \sim \;$}}
\def\gap{\lower.5ex\hbox{$\; \buildrel > \over \sim \;$}}

\input epsf
\begin{document}
\date{\today}
\title{Brane-World Creation and Black Holes}

\author{Jaume Garriga$^{1}$ and Misao Sasaki$^{2,3}$
\\
\hspace*{1cm}}

\address{$^{1}$IFAE, 
Departament de Fisica, Universitat Autonoma de Barcelona,\\
08193 Bellaterra $($Barcelona$)$, Spain$;$\\
$^{2}$Department of Earth and Space Science, Graduate School of Science,\\
 Osaka University, Toyonaka 560-0043, Japan$;$\\
$^{3}$Department of Physics, School of Science,\\
University of Tokyo, Tokyo 113-0033, Japan.
}

\maketitle

\thispagestyle{empty}
\vspace{1cm}
\centerline{\bf Abstract}
\vspace{2mm}
\begin{abstract}
An inflating brane-world can be created from ``nothing'' together with its 
Anti-de Sitter (AdS) bulk. The resulting space-time has compact spatial 
sections bounded by the brane. During inflation, the continuum of KK 
modes is separated from the massless zero mode by the gap $m=(3/2) H$, 
where $H$ is the Hubble rate. We consider the analog of the 
Nariai solution and argue that it describes pair production of 
``Black cigars'' attached to the inflating brane. In the case when the 
size of the instantons is much larger than the AdS radius, the 5-dimensional 
action agrees with the 4-dimensional one. Hence, the 5D and 4D gravitational 
entropies are the same in this limit. We also consider thermal instantons 
with an AdS black hole in the bulk. These may be interpreted as describing 
the creation of a hot universe from nothing, or the production of AdS black 
holes in the vicinity of a pre-existing inflating brane-world.
The Lorentzian evolution of the brane-world after
creation is briefly discussed. An additional "integration constant" in the
Friedmann equation -accompanying a term which dilutes like radiation- 
describes the tidal force in the fifth direction and arises from the mass 
of a spherical object inside the bulk. In general, this could be a 
5-dimensional black hole or a "parallel" brane-world of negative tension 
concentrical with our brane-world. In the case of thermal 
solutions, and in the spirit of the  $AdS/CFT$ 
correspondence, one may attribute the additional term to
thermal radiation in the boundary theory. Then, for temperatures 
well below the AdS scale, the entropy of this radiation agrees 
with the entropy of the black hole in the AdS bulk.
\end{abstract}

\hspace*{1cm}
$~~~~~~~~~~~~~~~~~~~~~~~~~~~~~~~~~~~~~~~~~~~~~~~~~~~~$ OU-TAP 108
$~~~~$UAB-FT 479 
\vspace{1cm}
\newpage

\section{Introduction}

The idea that we may live on a brane propagating in a bulk spacetime of
higher dimension is currently being considered in a variety of contexts,
ranging from M-theory \cite{strings} to phenomenological particle 
physics \cite{gia}. In particular, Randall and Sundrum (RS)
have proposed a simple and attractive
scenario where our brane-world is embedded in a 5-dimensional bulk with a 
negative cosmological constant $\Lambda_5$ \cite{RS1,RS2}. 
In this scenario, the extra dimension need not be small. In fact it could be 
infinite and yet we would perceive gravity as effectively 4-dimensional 
\cite{RS2,SMS,GT}. The reason is that the bulk is strongly curved,
so that most of its physical volume is within
a short distance $\ell\sim \Lambda_5^{-1/2}$ from the brane.
The picture of a brane-world evolving in a larger spacetime brings
with it a new perspective for early universe cosmology. In particular,
it seems natural to investigate the question of brane-world creation. 

The dynamics of the gravitational field on the RS brane 
has been recently analyzed in \cite{SMS,GT}. For the case of a single 
brane with positive tension, it has been shown that
4D Einstein gravity is recovered on the wall, with some corrections
at short distances and at high matter densities \cite{SMS,GT}. For the 
case of two parallel branes with opposite tension, there are additional 
scalar interactions corresponding to the brane separation modulus.
These give rise to Brans-Dicke type theories on the branes with 
rather interesting behaviour \cite{GT} (the case where the dilaton
is stabilized has received some attention in \cite{rst}). 
A discussion of gravitational collapse
of matter on the brane was given in \cite{CHR}. 
It was argued that the resulting "black hole" would be a 
5-dimensional object attached to the brane, with an event horizon in the 
shape of a cigar. This conjecture was confirmed in the analogous 2+1 
dimensional case \cite{EHM}. In the 3+1 dimensional case, the conjecture 
is consistent with the form of the weak field created by spherical matter 
sources on the wall \cite{GT}, although the non-perturbative "black cigar"
solution has not yet been found. Finally, a great deal of attention has
been devoted to cosmological solutions \cite{cosmo,cosmo2}.
In any discussion of gravity on the 
brane, the issue of boundary conditions in the bulk is quite important. 
If we are interested in the gravitational field created by matter sources on
an asymptotically flat brane,
it seems reasonable to impose that the bulk is asymptotically AdS 
in the past Cauchy horizon and at spacelike infinity, 
and that there is no incoming graviton 
flux from the past Cauchy horizon \cite{GT}. 
In the cosmological context, however,
it seems much more appropriate to address the question of boundary 
conditions in the framework of quantum cosmology. 

In this paper, we shall consider the quantum creation of an inflating 
brane-world and its subsequent evolution. In Section II we shall present 
the de Sitter-brane instanton. In Section III we shall generalize it to 
find the analog of the Nariai solution. As we shall see, this solution 
can be interpreted as describing pair creation of black cigars attached to
the inflating brane. In Section IV we shall consider "thermal"
instantons which 
contain AdS black holes in the bulk. These may describe the creation of
a hot universe from nothing, or the production of AdS black holes in the
vicinity of a preexisting inflating brane-world. Our conclusions are 
summarized in Section V.

\section{de Sitter-brane instanton}

In 4-dimensional gravity, the simplest description of the
birth of the universe involves the de Sitter instanton. This is a
4-sphere which interpolates between a point (or "nothing") at the south
pole, and a 3-sphere of maximal radius at the equator. The 3-sphere 
represents the geometry of the universe at the moment of creation. The 
subsequent cosmological evolution is given by the analytic continuation 
of the 4-sphere to Lorentzian signature, which gives an inflating 
spacetime. 

This picture can be generalized to the case of a brane-world embedded 
in an AdS bulk.
The line element of a 5-dimensional Euclidean AdS space can be
written as
\begin{equation}
ds_E^2= g_{ab}dx^a dx^b= dr^2+\ell^2 \sinh^2(r/\ell)[d\chi^2+\sin^2\chi 
d\Omega_{(3)}^2].
\label{ads1}
\end{equation}
Here $d\Omega_{(3)}^2$ is the metric on the 3-sphere and 
$\ell=(-6/\Lambda_5)^{1/2}$ is the AdS radius. A compact brane-world 
instanton can be constructed by excising the spacetime region at $r> r_0$ 
and gluing two copies of the remaining spacetime along the 4-sphere 
at $r=r_0$. On that 
hypersurface a brane of tension $\sigma$ is introduced so that Israel's
matching conditions \cite{SMS}
\begin{equation}
\partial_r g_{\mu\nu}|_{r=r_0} = {8\pi G_5\over 3}\, 
\sigma g_{\mu\nu}|_{r=r_0},
\label{israel}
\end{equation}
are satisfied. Here 
$G_5$ is the 5-dimensional gravitational constant and
\begin{equation}
\sigma={3\over 4\pi G_5\ell}\coth(r_0/\ell).
\label{sigmadef}
\end{equation}
The greek indices run over the coordinates of the 4-sphere.
The result of this cut and paste procedure is illustrated in Fig.~1. 
Two copies of a spherical patch of AdS bulk are bounded by a common 
4-sphere, which is the world-sheet of the brane. We may 
further identify both copies of the bulk by imposing a $Z_2$ symmetry,
although this identification will not play a prominent role in our
discussion.

\begin{figure}[t]
\centering
\hspace*{-4mm}
\epsfysize=10 cm \epsfbox{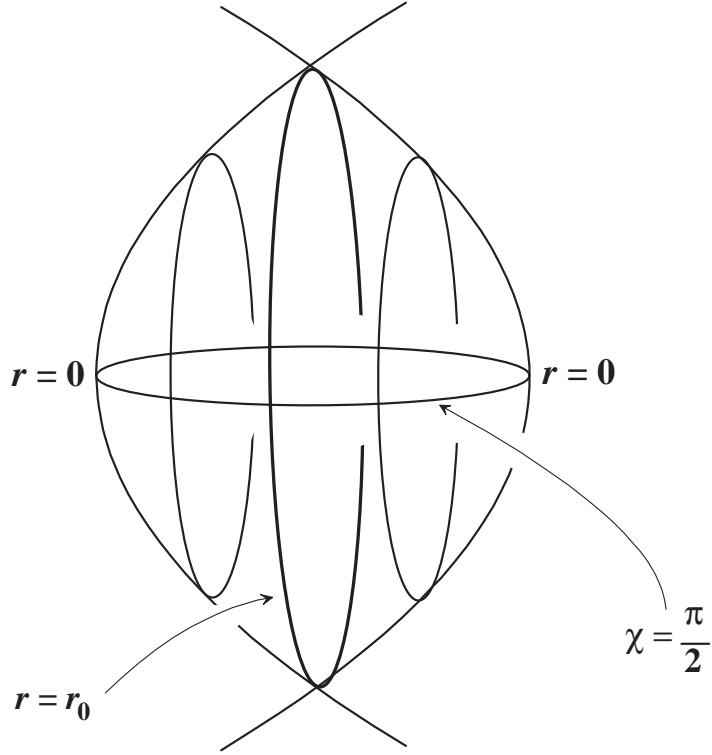}\\[3mm]
\label{fig1}
\caption[fig1]{De Sitter-brane instanton. The thick vertical
circle at $r=r_0$ represents the 4-sphere brane at which
the two identical 5-dimensional anti-de Sitter spaces are glued.
The $\chi=\pi/2$ hypersurface is the nucleation geometry.
}
\end{figure}
\vspace*{-5mm}
\vspace*{5mm}

The instanton in Fig.~1 can be interpreted as a semiclassical path
for the creation of a universe from nothing. Cutting the instanton 
in half, the Euclidean solution interpolates between ``nothing'', at 
the south pole, and a spherical brane of radius
$$
H^{-1}\equiv \ell \sinh(r_0/\ell)
$$
at the equator. This is completely analogous to
the usual 4-dimensional de Sitter instanton, except that in the
present case the inside of the brane-world is filled with AdS bulk.

The evolution of the brane after creation is given by the analytic
continuation of (\ref{ads1}) to real time. This continuation is done
by means of the substitution $\chi\to i H t + (\pi/2)$, which leads to
\begin{equation}
ds^2= dr^2+ (\ell H)^2 \sinh^2(r/\ell)
[-dt^2+H^{-2}\cosh^2(Ht) d\Omega_{(3)}^2]. \quad (r\leq r_0)
\label{ads2}
\end{equation}
The Lorentzian spacetime is represented in Fig.~2.

\begin{figure}[t]
\centering
\hspace*{-4mm}
\epsfysize=10 cm \epsfbox{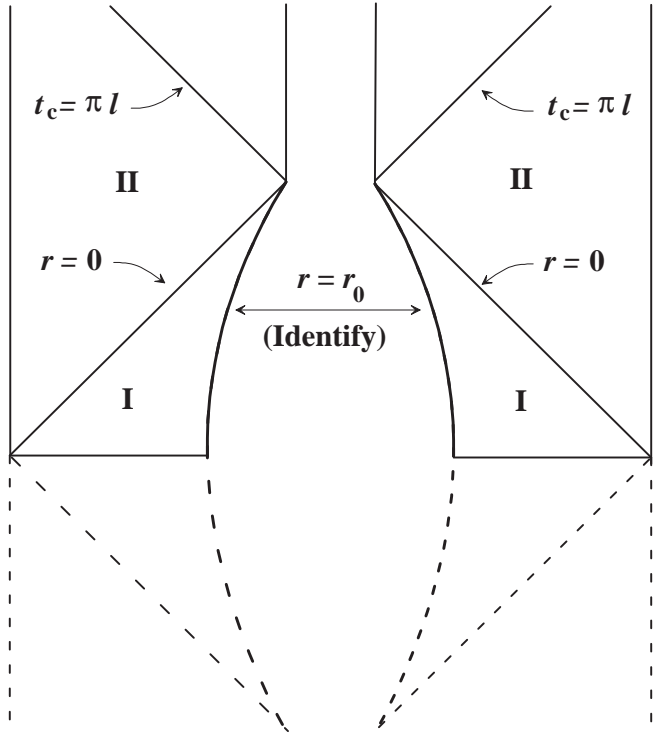}\\[3mm]
\label{fig2}
\caption[fig2]{Conformal diagram of the inflating brane-world
(at $r=r_0$) created from nothing. Each point in the figure represents a
3-sphere.} 
\end{figure}
\vspace*{-5mm}
\vspace*{5mm}

The brane at $r=r_0$ is now an inflating de Sitter space with Hubble rate 
$H^{-1}$. The line element (\ref{ads2}) does not cover the
whole spacetime, but only the exterior of the Rindler horizon
at $r=0$ (region I). This horizon is the light-cone emanating from the
center of symmetry of the solution.
To cover the interior of the light-cone (region II), we can make
the complexification $r \to i t_c$, $H t \to r_c -i (\pi/2)$.
In region II, the metric is that of a 5-dimensional open FRW universe,
$$
ds^2= -dt_c^2+ \ell^2 \sin^2 (t_c/\ell)[dr_c^2+\sinh^2 r_c d\Omega_{(3)}^2]
$$
where $r_c \sim 1$ corresponds
to the scale of spatial curvature. The coordinate $t_c$ runs from $0$
to $\pi\ell$. At $t_c=\pi\ell$ the open universe ``recollapses''. The null 
surface $t_c=\pi\ell$ is a Cauchy horizon accross which we can match 
our solution to a new AdS region or to a new de Sitter-brane region. 
But in any case, this extended solution is just a mathematical idealization, 
too simple to represent the real world.

Inflation will only be useful if it comes to an end, giving rise 
to thermalized regions. For that reason, we should consider the situation
where the effective tension $\sigma$ is partly due to the potential energy 
of an inflaton field $\varphi$:
\begin{equation}
\sigma=\sigma_c + V(\varphi).
\label{sigmaeff}
\end{equation}
Here, we have arbitrarily separated from $V$ the constant part,
\begin{equation}
\sigma_c = {3\over 4\pi \ell G_5}.
\label{sc}
\end{equation}
A brane which has the ``critical'' tension $\sigma_c$ does not
inflate in the AdS background (the static
Randall-Sundrum flat brane-world \cite{RS2} can be recovered from 
(\ref{ads2}) in the limit $r_0\to\infty$, which from (\ref{sigmadef})
corresponds to $\sigma=\sigma_c$), so the effective cosmological 
constant vanishes for $V=0$.

The cosmological evolution
of the closed universe after creation can be obtained as follows.
Due to spherical symmetry, the bulk inside a closed brane-world will
be simply AdS, which is conveniently described in the static
chart as
\begin{equation}
ds^2=-A(R)dT^2+{dR^2\over A(R)}+R^2d\Omega^2_{(3)}\,,
\label{static}
\end{equation}
where $A(R)=(1+R^2/\ell^2)$. The brane-world trajectory can be
parametrized as $R=R(t)$, 
$T=T(t)$. There is much freedom in this parametrization, and we shall
choose the gauge in which 
\begin{equation}
\dot R^2=A(R)^2 \dot T^2 -A(R).
\label{protof}
\end{equation}
With this choice, the metric on the brane is given by
$$
ds_4^2=-dt^2+R^2(t) d\Omega^2_{(3)}.
$$
The equation of motion for the scale factor $R$ can be obtained from
Israel's matching conditions under the assumption of $Z_2$ symmetry: 
\begin{equation}
K_{\mu\nu}=-4\pi G_5[T_{\mu\nu}-
(1/3)T g_{\mu\nu}], \label{extcon}
\end{equation}
where $T_{\mu\nu}$ is the energy momentum tensor on the brane and 
$K_{\mu\nu}$ is the extrinsic curvature. In terms of the normal vector
$n_a=(\dot R,-\dot T,0,0,0)$, this is given by
$K_{ab}=-(g_a^c-n_an^c)(g_b^d-n_bn^d)n_{c;d}$. The angular
components are given by 
$$
K_{\Omega\Omega'}=n_a\Gamma^a_{\Omega\Omega'}
=-\dot T [A(R)/R] g_{\Omega\Omega'}.
$$
Assuming that the energy momentum tensor has the perfect fluid form
$$
T^{\mu}_{\nu}=\hbox{diag}(-\rho,p,p,p)-\sigma_c \delta^{\mu}_{\nu},
$$
where $\sigma_c$ is given in (\ref{sc}) and defining
$$
X={8\pi G\over 3} \rho,
$$
the angular components of (\ref{extcon}) give
\begin{equation}
\left({\dot R\over R}\right)^2={1\over \ell^2}-{A(R)\over R^2}+X+
{\ell^2\over 4} X^2.
\label{frwgen}
\end{equation}
It can be shown that the temporal components of (\ref{extcon}) are in
fact redundant. With $A(R)=1+R^2/\ell^2$ we have
\begin{equation}
\left({\dot R\over R}\right)^2=-{1\over R^2}+{8\pi G\over 3} \rho +
\ell^2\left({4\pi G\over 3} \rho\right)^2.
\label{modfrw}
\end{equation}
Hence, the standard Friedmann equation for cosmological evolution is
recovered \cite{note}, with a correction term which is negligible
at moderately low energy densities $\ell^2 X \ll 1$. 

Returning our attention to the inflating de Sitter solution, let us now 
check the existence of the bound state of gravity on the
brane and the nature of the spectrum of Kaluza-Klein (KK) excitations.
Expanding the metric as $ds^2=(g_{ab}+h_{ab})dx^a dx^b$, 
we shall work in the analog of the Randall-Sundrum (RS) gauge,
\begin{equation}
h_{rr}=h_{r\mu}=h^{\mu}_{\mu}=h^{\mu\nu}{}_{|\nu}=0,
\label{rsgauge}
\end{equation}
where the vertical stroke indicates covariant derivative in the 
4-dimensional de Sitter space.
It is convenient to define $\hat h_{\mu \nu} = a^{-1/2} h_{\mu\nu}$, where
$a(r)=\ell \sinh(r/\ell)$, and a conformal radial coordinate $\eta$
through $dr=a\ d\eta$. In terms of $\eta$ we have
$$
a(\eta) = {\ell\over \sinh(|\eta|+\eta_0)}, \quad (-\infty < \eta < +\infty) 
$$
where $\eta_0$ is defined by the equation $\sinh \eta_0= 1/\sinh(r_0/\ell)$.
The equation of motion for perturbations in
the gauge (\ref{rsgauge}) can be separated as \cite{KS}
\begin{equation}
-\hat h_{\mu\nu}''+
{(a^{3/2})''\over a^{3/2}} \hat h_{\mu\nu}
= {m^2\over H^2} \hat h_{\mu\nu},
\label{schrodinger}
\end{equation}
\begin{equation}
(-\Box + 2H^2 + m^2) \hat h_{\mu\nu}=0.
\label{graviton}
\end{equation} 
The Schrodinger equation (\ref{schrodinger}) determines the spectrum of 
masses $m^2$, while (\ref{graviton}) is the equation of motion for spin
two fields of mass $m^2$ in a de Sitter space of radius $H^{-1}$. The
box indicates the 4-dimensional covariant d'Alembertian. 

Equation (\ref{schrodinger}) has the obvious normalizable 
bound state $\hat h_{\mu\nu}\propto a^{3/2}$, which corresponds to 
the massless graviton. However, it is clear that the ``volcano potential'' 
in Eq. (\ref{schrodinger}) approaches the constant $9/4$ at 
$\eta\to \pm \infty$. Hence, the continuous spectrum of KK excitations
starts at 
$$
m=(3/2) H.
$$ 
This is a somewhat special value of the mass
in de Sitter space, corresponding to the critically damped case. For
smaller mass, long wavelenth perturbations are underdamped,
whereas for larger mass they have oscillatory behaviour.
As is well known, de Sitter space behaves in some respects as a
system with temperature $T=H/2\pi$. The mass gap between the zero mode 
and the KK modes may be related to the stability of de inflating 
brane-world against thermal excitation of the massive modes. 
This issue deserves further investigation.

Solutions with two concentrical branes can easily be constructed by
inserting a new brane of negative tension 
\begin{equation}
\tilde\sigma= -{3\over 4\pi G_5\ell}\coth(r_1/\ell) < 0
\label{negten}
\end{equation}
at some radius $r_1<r_0$. The region $r>r_1$ is excised,
and the edges of the two copies of the bulk are again identified along
the 4-sphere at $r_1$. In the absence of a mechanism stabilizing
the interbrane distance, the subsequent cosmological evolution
on both branes will be independent (although the evolution of cosmological 
perturbations will not be so, in general). In particular, the branes 
will in general drift away or come closer together in the course of 
evolution.

An important property of our scenario is that the problem of boundary
conditions on the past cauchy horizon does not exist. Our brane-world is
entirely contained in the causal future of the compact 
hypersurface at which the whole universe appears in the Lorentzian
regime. Hence the evolution of the brane-world is uniquely 
determined by the initial data.
One can then unambiguously calculate the quantum fluctuations around
the classical solution. 
One can either adopt Hartle-Hawking's no-boundary
boundary condition \cite{HH}, or Vilenkin's tunneling boundary
condition \cite{Vil}.
In the 4-dimensional theory, it is known that both boundary conditions 
give the same fluctuation spectrum \cite{GaV}.
In the present 5-dimensional theory,
detailed evaluation of the quantum fluctuation spectrum is left for
future study, but the analysis of the KK modes given above indicates
that only the massless (graviton) mode contributes. Hence the
result would be similar to the case of standard
4-dimensional quantum cosmological inflation, irrespective of
the Hartle-Hawking or Vilenkin boundary conditions.

\section{Pair production of "black cigars"}

As in the case of flat brane-worlds, it is possible to find nonlinear
generalizations of the zero mode on the brane. With the 
ansatz,
\begin{equation}
ds^2=g_{ab} dx^a dx^b = dr^2+ a^2(r) \gamma_{\mu\nu} dx^{\mu}dx^{\nu},
\label{gen}
\end{equation}
where $\gamma_{\mu\nu}$ is the 4-dimensional metric and 
$a(r)=\ell \sinh(r/\ell)$,
the 5-dimensional Riemann tensor is given by 
\begin{equation}
R^{\mu}{}_{r\nu r}= -\ell^{-2} \delta^{\mu}_{\nu}, \quad
R^{\mu}{}_{\nu\rho\sigma}= {}^{(\gamma)}R^{\mu}{}_{\nu\rho\sigma}
+ \cosh^2(r/\ell) (\delta^{\mu}_{\sigma} \gamma_{\nu\rho} -
\delta^{\mu}_{\rho} \gamma_{\nu\sigma}).
\label{riemann}
\end{equation}
It is then straightforward to show 
that the 5-dimensional equations of motion 
$R_{ab}= -4 \ell^{-2} g_{ab}$ are satisfied provided that
$\gamma_{\mu\nu}$ is a solution of
\begin{equation}
{}^{(\gamma)}R_{\mu\nu}= 3 \gamma_{\mu\nu}, 
\label{con}
\end{equation}
Brane solutions can be constructed in the same manner
as in the previous section, by excising the region with $r>r_1$ and 
introducing at the boundary a brane of tension given by (\ref{sigmadef}).
The metric on the brane will be given by $a^2(r_0)\gamma_{\mu\nu}$. 
The argument is valid both for Euclidean and for Lorentzian metrics.
Equation (\ref{con}) can be interpreted as
the 4-dimensional Einstein equations on the brane in the presence of
a positive cosmological constant 
$$
\Lambda_4= {3\over a^2(r_0)} \equiv 3H^2.
$$
One solution of (\ref{con}) is de Sitter space, but any other solution 
can be chosen. As in the previous section, we may optionally
introduce a second brane of negative tension
(\ref{negten}) at some $r_1<r_0$, excising the region with $r<r_1$.

The Euclidean action for the system of two branes of 
tension $\sigma_i$ located at $r_i$ and separated by a
region of AdS bulk is given by
\begin{equation}
S_E= {1\over 16\pi G_5} \int d^5x \sqrt{g}(2 \Lambda_5-{\cal R})+
\sum_i \sigma_i a^4(r_i)\int d^4x\sqrt{\gamma}
\label{eact}
\end{equation}
where ${\cal R}$ is the Ricci scalar. The energy momentum tensor is given
by
$$
T_{ab}=-(8\pi G_5)^{-1}\Lambda_5 g_{ab}-
\sum_i \sigma_i a^2 \gamma_{ab} \delta(r-r_i),
$$ 
where $\gamma_{ab}=0$ when $a$ or $b$ are equal to $r$. 
Using the trace of the Einstein 
equations, ${\cal R}$ can be elliminated from
(\ref{eact}), which gives
\begin{equation}
S_E=-{1\over 12\pi G_5}\int d^5 x\sqrt{g} \Lambda_5
-{1\over 3} \sum_i\sigma_i a^4(r_i)\int d^4 x \sqrt{\gamma}.
\end{equation}
Performing the integrals we have 
\begin{equation}
S_E= \sum_{i=0}^1 (-1)^i \left(\coth(r_i/\ell)
-{r_i/\ell\over\sinh^2(r_i/\ell)}\right)S_E^{(i)},
\label{actionsum}
\end{equation}
where
\begin{equation}
S_E^{(i)}=-{3 V_4^{(\gamma)} \over 8\pi G H_i^2}.
\label{action4}
\end{equation}
Here
$$
G=G_5/\ell
$$
is the 4-dimensional Newton constant (defined in the limit of
low energy world, or when the effective tension of the brane
is close to $\sigma_c$ and the matter density is much
lower than $\sigma_c$), $H_i=[\ell \sinh(r_i/\ell)]^{-1}$,
and $V_4^{(\gamma)}$ is the (dimensionless) volume of 
the manifold with metric $\gamma_{\mu\nu}$. In the case of a single 
brane of positive tension, only the first term in (\ref{actionsum}) is
present. Also, in the limit 
where $r_0\gg r_1, \ell$, we have 
$S_E\approx S_E^{(0)}$, which in turn reduces to the usual 
4-dimensional action of the instanton with metric 
$a^2(r_0)\gamma_{\mu\nu}$ and cosmological constant
$\Lambda_4= 3 H_0^2$.

It should be noted that the one-brane 5-dimensional
solutions of the form (\ref{gen}) will in general be singular. 
Indeed, from (\ref{riemann}), the non-vanishing components of the 
5-dimensional Weyl tensor are given by 
$C^{\mu}{}_{\nu\rho\sigma}= {}^{(\gamma)}C^{\mu}{}_{\nu\rho\sigma}$,
and therefore
$$
C^2=^{(\gamma)}C^2/a^4(r).
$$
Unless $^{(\gamma)}C^2=0$ (as is the case when the brane metric is
conformally flat) this invariant diverges in the vicinity of $r=0$. Of
course, this singularity will not occur in the case with two branes.
A similar divergence occurs in the Randall-Sundrum solution \cite{RS2}
when we substitute the flat 4-dimensional metric on the brane by any 
Ricci flat metric $\gamma_{\mu\nu}$
\begin{equation}
ds^2=dy^2+a^2(y)\sigma_{\mu\nu}dx^{\mu}dx^{\nu},
\end{equation}
where $a(y)=\ell e^{|y|/\ell}$. In this case the divergence is at the
AdS horizon $y\to \infty$. An exception occurs in the case when 
$\sigma_{\mu\nu}$ is
a 4-dimensional plane wave \cite{CG}. The reason is that plane waves
are null, and any invariant such as $^{(\sigma)}C^2$ vanishes. 
In the cosmological case, $^{(\gamma)}C^2$ does 
not vanish even for linearized gravity waves.  
This applies to the zero mode discussed in the previous section, 
which is therefore singular on the Rindler horizon. However, we do not
believe that this indicates any fundamental difficulty. The expansion in 
modes is just an instrument which is useful in order to find the Green's
function of the wave equation. For instance, in the case of flat slices 
discussed by Randall and Sundrum, the KK modes are singular on the 
AdS horizon, and yet the gravitational field created 
by isolated sources on the brane (which is computed by using the Green 
function of the wave operator) is non-singular \cite{GT,SSM2}.

Let us now consider the Schwarzschild-de Sitter solution on the brane:
\begin{equation}
H^{-2} d\sigma^2= - F(R) dT^2+{dR^2 \over F(R)}+R^2 d\Omega_{(2)}^2.
\label{sds}
\end{equation}
Here $F(R)= (1-H^2 R^2 - 2M/R)$ and $M$ is a free parameter. For
$HM\ll 1$, the equation $F(R)=0$ has three real solutions. One of 
them is negative and the other two are positive. The two positive roots 
correspond to the black hole and cosmological horizons respectively. 
For $HM\ll 1$ the 5-dimensional solution is analogous to the 
``black string'' described in \cite{CHR}. At large distances from 
the brane $\sinh(r/\ell)\ll(HM)^{-1}$, the physical size of the horizon 
$a(r) HM$ is much smaller than the 
AdS radius $\ell$. In this region, the 
horizon of the black string locally resembles a cylinder, which is
entropically unstable to fragment into an array of spherical 
horizons which in total have a larger surface area \cite{ruth}.

If the parameter $M$ in (\ref{sds}) is increased, the 
black hole horizon on the brane increases and the cosmological horizon 
decreases. The extremal case 
where both horizons have the same size is the Nariai solution, 
which has $M=(3\sqrt{3}H)^{-1}$ and $R=(\sqrt{3}H)^{-1}$. The 
$R,T$ coordinates
become inadequate in this case, since $F$ tends to zero. 
It is customary to introduce a small parameter $\epsilon$, defined by 
$27 M^2H^2=1-3\epsilon^2$, and new coordinates $(\lambda,\psi)$ through
$\lambda=\epsilon \sqrt{3} H T$, 
$\sqrt{3} H R=(1-\epsilon\cos\psi-\epsilon^2/6)$. 
The black hole horizon is now at $\psi=0$, whereas the cosmological 
horizon is at $\psi=\pi$ \cite{GP}. 
In the limit $\epsilon\to 0$, one finds the Nariai solution:
\begin{equation}
d\sigma^2={1\over 3}[-\sin^2\psi\ d\lambda^2+d\psi^2+ d\Omega_{(2)}^2].
\label{nariai}
\end{equation}
Near this extremal case, the 5-dimensional solution is qualitatively
different from the black string. In fact, it appears to be much closer 
to the black cigar postulated in Ref. \cite{CHR}. The potentially 
unstable regime $a(r) HM \ll \ell$ only arises for $r\ll \ell$, where 
the ``scale factor'' behaves linearly 
$a \propto r$. In this region, the horizon does not look 
like a cylinder at all. In fact, the spatial section of the 
horizon closes off smoothly at $r=0$, and looks 
like Fig. 1, except that now the vertical circles are 2-spheres
rather than 4-spheres.
Therefore, the usual argument that the horizon 
should be entropically unstable, breaking up into small spheres, cannot be 
applied. Some instability may still 
be expected near the tip of the cigar, at $r=0$, where $C^2$ diverges. 
We shall argue, however, that the singularity at $r=0$ is rather mild, 
and may in fact be smoothed out by quantum fluctuations. 
Note, in this connection, that even for small zero mode perturbations 
near the de Sitter-brane solution, the invariant $C^2$ diverges near 
$r=0$. Zero point fluctuations corresponding to these modes, however, 
have finite action and will inevitably occur. Similar fluctuations 
may be expected to smooth out the tip of the black cigar in the 
Nariai-brane solution.

As is well known, the 4-dimensional Nariai instanton can be Euclideanized 
by making the substitution $\lambda\to i \phi$ in (\ref{nariai}).
The corresponding Euclidean solution is the product of two 2-spheres 
$S^{(2)}\times S^{(2)}$, and in the limit when
$r_0\gg \ell$, its action is given by 
(\ref{action4}) with $V_4^{(\gamma)}=(4\pi/3)^2$ and $r_i=r_0$,
\begin{equation}
S_E^N\approx-{2\pi\over 3 GH^2}
\label{nariaiact}
\end{equation}
Interestingly, this is the same as the action of the 4-dimensional 
Nariai instanton with the same curvature radius. 
For compact instantons, $-S_E$ coincides
with the gravitational entropy. Hence, as long as the size of the instanton
is large compared with the ``cut-off'' scale $\ell$, the 
entropy of the 5-dimensional solution coincides with the entropy of the 
brane-world at the boundary. This is actually true for any choice of
the brane-world geometry, and is perhaps the consequence of a deeper
holographic principle. 

As mentioned above, the 5-dimensional instanton
is singular at $r=0$. However, by introducing 
a boundary at some small $r$, the Hawking-Gibbons boundary term, 
$$
-{a^4 \over 8\pi G_5} \int \sqrt{\gamma}
K^{\mu}_{\mu} d^4x= {3 a^3 \partial_r a\over 16 \pi G_5}V_4^{(\gamma)}
$$
vanishes as $r\to 0$ (here $K$ is the extrinsic curvature). 
Hence, this boundary does not contribute to the
action. Since the singularity is so mild, we believe that the
Nariai-brane instanton indeed represents a legitimate saddle
point giving the main contribution to the pair-creation rate of
"black cigars" attached to the brane-world. This rate is obtained
by exponentiating the difference between (\ref{nariaiact}) and
the action for the de Sitter-brane solution, 
$$
S_E^{dS}\approx-{\pi\over GH^2}.
$$
This gives the usual 4-dimensional result \cite{GP} 
$\Gamma\sim \exp(-\pi/3GH^2)$.

In an alternative scenario, the singularity can be avoided by introducing a 
membrane of negative tension, as mentioned at the beginning of this 
section. For sufficiently large absolute value of the tension, the 
contribution of this second brane-world to the action will be negligible
and we recover (\ref{nariaiact}). 

Finally, we may speculate about the existence of additional instantons 
describing the pair creation of black holes during inflation. 
One reason to suspect the existence of such instantons is that the 
black cigar on a flat brane has some corrections with respect to the
Schwarzschild solution, whereas the instanton presented here has no
such corrections with respect to the Nariai solution.
However, it should be recalled that the mass of the
black holes considered here is not arbitrary: it is related to 
the curvature of de Sitter and some ``miraculous'' cancellation of 
the corrections may occur for this special value of the mass. 
Nevertheless we should leave the door open to the possible existence 
of completely non-singular solutions. If these existed, they may have 
slightly lower action than the ones we have considered here. However, 
even in this case we expect that their action would be very similar in 
the low energy limit (since in this limit the action of our instanton 
agrees with the four-dimensional Nariai action).

\section{AdS black holes and thermal instantons}

In this section we shall consider instanton solutions
where the black hole is not on the brane, but in the bulk.
We assume a scalar field as the matter on the brane.
Since the instanton is compact, we may interpret these configurations 
as describing the birth of a brane-world which contains a bulk black
hole. An alternative interpretation is that they give the probability
for "pair" creation of bulk black holes inside of a pre-existing
inflating brane-world, analogous to the creation of black holes in the
presence of a domain wall in four dimensions \cite{CCG,note,BC,GenS}.
This interpretation parallels the one usually attributed to the 
4D Nariai solution.

The most general form for the spherically symmetric Euclidean space in
the bulk is the Schwarzschild-Ads solution, given by
\begin{equation}
ds^2=A(R)dT_E^2+{dR^2\over A(R)}+R^2d\Omega^2_{(3)}.
\label{static2}
\end{equation}
Here 
$$
A(R)=1+{R^2\over \ell^2}-{\alpha^2\over R^2},
$$
where the parameter $\alpha$ is related to the mass of the five
dimensional black hole. A spherial brane can be introduced, 
with trajectory given by $R=R(t_E)$, $T=T(t_E)$. Two identical
copies of the bulk bounded by this trajectory are glued at the brane.
Choosing the
parameter $t_E$ to be the cosmological proper "time" on the 
brane, the equations of motion reduce to the Euclidean version
of Eq. (\ref{modfrw}),
\begin{equation}
-\left({\dot R\over R}\right)^2=-{1\over R^2}+{\alpha^2\over R^4}+
\left(X_E+{\ell^2\over 4}X^2_E\right).
\label{emodfrw}
\end{equation}
Here $X_E=(8\pi G/3) \rho_E$, with 
$\rho_E=V(\phi)-(1/2)\dot \phi^2$ and the dot indicates derivative
with respect to $t_E$. This equation should be supplemented with 
the equation of motion for the scalar field,
\begin{equation}
\ddot \phi + 3{\dot R\over R} \dot \phi = V'(\phi).
\label{efield}
\end{equation}
Taking the derivative of (\ref{emodfrw}) and using (\ref{efield})
we obtain
$$
-\ddot R = {-\alpha^2\over R^3}
+R\left(X_E+{\ell^2\over 4} X_E^2\right)
+ 4\pi G \dot\phi^2 R \left( 1+ {\ell^2\over 2} X_E\right).
$$
Let $\phi_0$ be an extremum of $V(\phi)$. Then, it is straightforward
to show that there is a static solution with $\phi=\phi_0$ and
\begin{equation}
R_0^2= 2 \alpha^2 = [2(X_0+\ell^2X_0^2/4)]^{-1}.
\label{r0}
\end{equation}
The effective tension of the brane is given by $\sigma=\sigma_c+V(\phi_0)$.
Note that both the radius of the brane and the mass parameter $\alpha$
are given in terms of the energy density on the brane at the stationary
point.

The bulk metric (\ref{static2}) has a conical singularity in the 
$(R,T_E)$ plane at $R=0$, unless the the range of the "angular"
variable $T_E$ is adjusted as
$$
0<T_E\leq \beta = 4\pi \lim_{R\to R_+} {(R-R_+)\over A(R)}.
$$
Here 
$$
R_+^2={1\over 2} \left(\ell \sqrt{\ell^2+4\alpha^2} -\ell^2\right)
$$
is the black hole horizon radius. 
The temperature on the brane will be given by the inverse of 
the periodicity of the proper time, 
\begin{equation}
T_{b}= \beta^{-1} A^{-1/2}(R_0)= {1\over\sqrt{2}\pi R_+}.
\label{temperature}
\end{equation}
Hence the temperature felt by observers on the brane 
is proportional to the inverse of the black hole radius.

The Euclidean action of the thermal instanton can be calculated
along the same lines discussed in Section II, and it is given 
by
\begin{equation}
S_E=-{\pi^2 R_+^3\over G\ell}.
\end{equation}
Note that this expression also follows from the area law. 
{}From the 
five-dimensional point of view, the entropy of this system is 
due to the area of the AdS black hole, which is equal to $4 \pi^2 R_+^3$
(two copies of the black hole must be considered). Since
$G_5=G\ell$, we have 
\begin{equation}
S_{bh}=-S_E= {\rm Area}/4G_5.
\label{entropy}
\end{equation}
Incidentally, the same argument can be applied to the de Sitter and
Nariai instantons considered in the previous sections. There, the
action is equal to minus one fourth of the total area of cosmological
and black hole horizons, in natural units. 

For $\alpha\gg \ell$, the black hole radius, $R_+^2\approx \alpha\ell$,
is much larger than the AdS radius. In that case, 
the action of this instanton is much larger (smaller in absolute
value) than the action of the Nariai-brane instanton with the
same value of the effective brane tension (or 4-dimensional
cosmological constant). Hence, the nucleation rate of bulk black 
holes would be very suppressed compared with the nucleation rate
of black cigars on the brane.

On the other hand, if the thermal instanton is interpreted as
describing creation of a hot universe from nothing, it is of some 
interest to consider the evolution of this universe after creation. This
is described by the analytic continuation of (\ref{emodfrw}),
\begin{equation}
\left({\dot R\over R}\right)^2={-1\over R^2}+
\left({8\pi G\over 3}\right) \rho
+ \ell^2 \left({4\pi G\over 3}\rho\right)^2+
{\alpha^2\over R^4},
\label{bhfrw}
\end{equation}
where $\rho$ is the density of matter on the brane.
This is very similar to the usual Friedmann equation for the evolution
of a 4-dimensional universe in Einstein gravity. The only differences
are the third and fourth terms in the right hand side of the equation.
The third term is the correction at large densities which we had already 
encountered in Section II. The
last term, which is due to the presence of the black hole in the bulk,
behaves from the point of view of four dimensions as an
additional contribution to the matter density which dilutes like radiation. 

The possible presence of an additional term 
which dilutes like radiation was noted in Ref. \cite{cosmo2}
for the case of spatially flat cosmology. There, the 
analog of $\alpha$ appeared as an integration constant
of the equations of motion on the brane. As shown above, 
this term corresponds to the existence of a black hole in the bulk
\cite{note,Kra,ida}.
In fact, any spherically symmetric massive object in the bulk 
(or even a "massive" parallel brane-world concentrical with ours) would 
cause the same effect. This additional term does not arise in 
the case of the maximally symmetric de Sitter-brane instanton. In principle 
it would be present if 
the universe was born via the 
thermal instanton. However, since the universe must 
inflate after creation, it seems that in the post-inflationary epoch the 
most reasonable choice for the "integration constant" is zero.

Recently, there has been much interest in the so-called $AdS/CFT$ 
correspondence, a possible duality relation 
between classical gravity in a five dimensional AdS bulk and a 
conformal field theory living on its boundary \cite{gub}. In this context, 
the last term in Eq. (\ref{bhfrw}) may be interpreted from the boundary 
theory point of view, as the contribution of a true radiation bath of the 
conformal fields whose energy 
density is $\rho_r=(3/8\pi G)(\alpha^2/R_0^4)$, whose temperature is 
$T_b$ and whose volume is given by $2\pi^2 R_0^3$. The entropy of such
radiation bath is given by 
$$
S_{rad}={\pi^2\alpha R_{+}\over G}.
$$
We note that for $\alpha\gg\ell$, this coincides with the entropy of 
the 5D bulk black hole given by Eq. (\ref{entropy}),
$$
S_{rad}(4D)\approx S_{bh}(5D),
$$
in agreement with the duality conjecture.

\section{conclusions}

We have presented a quantum cosmological scenario of brane-world creation.
In this scenario, not only an inflating brane but also the associated
5-dimensional bulk spacetime are created from nothing. The
quantum creation is described by the de Sitter-brane instanton which
consists of two identical patches of Anti-de Sitter space glued
together by an $S^{(4)}$ brane. The Euclidean action of the instanton
is generally different from that of the 4-dimensional de Sitter
instanton but reduces to it if the de Sitter radius is large compared with
the AdS curvature radius. We have found that the gravitational
Kaluza-Klein modes have a mass gap in the spectrum and the continuous
spectrum starts above the mass $m=(3/2)H$, where $H$ is the expansion
rate during inflation. 

We have also found that there is a family of 5-dimensional solutions 
such that the metric on the brane is any solution of the 4-dimensional vacuum 
Einstein equations with a cosmological constant. Unless the 4D metric is
conformally flat, these solutions are singular at the center of the AdS bulk.
But the singularity is so weak  that it does not affect the value of the
action at all (the singularity can also be avoided by placing a negative 
tension brane at a small radius without altering the Euclidean action much.)
Among these solutions is the Nariai-brane instanton which has the topology of
$S^{(2)}\times S^{(2)}$ on the brane. We have argued that this instanton
describes pair-creation of black cigars threading through the brane.

We have studied thermal instantons which contain an AdS black hole
in the bulk. These may either represent the creation of a hot universe 
from nothing or the pair creation of AdS black holes near an inflating
brane. In the latter interpretation, however, the nucleation 
rate of AdS black holes would be very much suppressed compared with
that of black cigars attached to the brane when the size of the
instantons is larger than the AdS radius. The tidal forces exerted 
by such black holes on the brane manifest themselves as an additional 
term in the Friedmann equation which dilutes like radiation.
In the spirit of the $AdS/CFT$ correspondence, this may be attributed
to thermal radiation in the boundary theory. In this case, for temperatures
much lower than the AdS scale, the entropy of this radiation agrees
with the entropy of the black hole in the five dimensional theory.

The fact that the Euclidean action of brane instantons
reduces to the corresponding 4-dimensional action in the low energy
limit is reassuring. In particular, this 
suggests that most of the predictions made in the standard 
4-dimensional quantum cosmological context are qualitatively
valid also in brane-world scenarios. An exception may occur
for instantons describing the creation of
open universes from nothing \cite{HT}. These instantons are 
singular in four dimensions, and therefore a different behaviour of
gravity at high densities and short distances can make a
difference in the Euclidean action. This case is left for further
study.

\section*{Acknowledgements}

We thank Tetsuya Shiromizu for useful discussions.
J.G. acknowledges support from CICYT under grant
AEN98-1093. M.S. 
acknowledges support from Monbusho, under Grant-in-Aid for Scientific
Research No. 09640355.
J.G. thanks X. Montes and T. Tanaka for useful discussions, and
the organizers of the 4th RESCEU symposium on 
``The Birth and Evolution of the Universe'' for their kind 
hospitality during the initial stages of this work.


\begin{thebibliography}{99}
\bibitem{strings} see, e.g., J. Polchinski, {\it String Theory I \& II}
(Cambridge University Press, Cambridge, 1998).
\bibitem{gia} N. Arkani-Hamed, S. Dimopoulos, G. Dvali,
Phys.Lett. B429 (1998) 263-272; Phys.Rev. D59 (1999) 086004;
I. Antoniadis, N. Arkani-Hamed, S. Dimopoulos, G. Dvali,
Phys.Lett. B436 (1998) 257-263; A. Pomarol, hep-ph/9911294.
\bibitem{RS1} L.~Randall and R.~Sundrum, Phys. Rev. Lett.{\bf 83}, 
3370 (1999).
\bibitem{RS2} L.~Randall and R.~Sundrum, hep-ph/9906064.
\bibitem{SMS} T.~Shiromizu, K.~Maeda and M.~Sasaki, 
gr-qc/9910076.
\bibitem{GT} J. Garriga and T. Tanaka, hep-th/9911055.
\bibitem{rst} C. Csaki, M. Graesser, L. Randall and J. Terning,
 hep-ph/9911406; W. Goldberger and M. Wise, Hep-th/9911457. 
\bibitem{CHR} A.~Chamblin, S.W.~Hawking and H.S.~Reall, 
hep-th/9909205.
\bibitem{EHM} R.~Emparan, G.T.~Horowitz and 
R.C.~Myers, hep-th/9911043.
\bibitem{cosmo} N. Kaloper, Phys.Rev. D60 (1999) 123506.
\bibitem{cosmo2}
P.~Bin\'etruy, C.~Deffayet, U.~Ellwanger and 
D.~Langlois, hep-th/9910219.
E.E. Flanagan, S.-H.H. Tye and I. Wasserman, hep-ph/9910498.
\bibitem{note} We were informed by T. Shiromizu that the Lorentzian
  version of a family of spherically symmetric brane solutions has been
  recently discussed by Kraus \cite{Kra} (see also \cite{ida}). 
\bibitem{KS} H. Kodama and M. Sasaki, Prog. Theor. Phys. Suppl.
{\bf 78}, 1 (1984).
\bibitem{HH} J.B. Hartle and S.W. Hawking, Phys. Rev. D {\bf 28},
2960 (1983).
\bibitem{Vil} A. Vilenkin, Phys. Rev. D {\bf 30}, 509 (1984);
Phys. Rev. D {\bf 33}, 3560 (1986).
\bibitem{GaV} J. Garriga and A. Vilenkin,
Phys. Rev. D {\bf 56} 2464 (1997).
\bibitem{CG} A.~Chamblin and G.W.~Gibbons, 
hep-th/9909130.
\bibitem{SSM2} M. Sasaki, T. Shiromizu and K. Maeda, hep-th/9912233.
\bibitem{ruth} R. Gregory and R. Laflamme, Phys. Rev. Lett. {\bf 70},
2837 (1993).
\bibitem{GP} P. Ginsparg and M.J. Perry, Nucl. Phys. B {\bf 222},
245 (1983).
\bibitem{CCG} R. Caldwell, A. Chamblin and G. Gibbons, Phys. Rev. D
{\bf 53}, 7103 (1996); R.B. Mann, Nucl. Phys. {\bf B516}, 357 (1998);
A. Chamblin and
J.A.M. Ashbourn-Chamblin, Phys. Rev. {\bf D57} 3529 (1998).
\bibitem{note} The use of compact Euclidean 
solutions in order to describe nucleation processes in a preexisting
background is the subject of some debate (see e.g. \cite{BC,GenS} for a 
recent discussion). The problem is that the instanton itself does not
represent a semiclassical path interpolating between the original
configuration and the configuration after tunneling. (incidentally,
this is true even for the well-known Coleman-de Lucchia instanton 
describing false vacuum decay in the presence of gravity). Here, we 
shall not dwelve into this problem, adopting the interpretation
given in the original references. Our emphasis will be in the case 
corresponding to creation from nothing, where the interpretation of 
instantons as semiclassical paths is perhaps less controversial.
\bibitem{BC}  R. Bousso and A. Chamblin, 
Phys. Rev. {\bf D59},084004 (1999). 
\bibitem{GenS} U. Gen and M. Sasaki, gr-qc/9912096, to be
published in Phys. Rev. D.
\bibitem{Kra} P. Kraus, hep-th/9910149.
\bibitem{ida} M. Cvetic, S.Griffies and H.H. Soleng,
Phys. Rev. {\bf D48}, 2613 (1993); M. Cvetic and H.H. Soleng
Phys. Rept. {\bf 282}, 159 (1997); A. Chamblin and H. Reall,
hep-th/9903225; D. Ida, gr-qc/9912002.
\bibitem{gub} See e.g. S.S. Gubser hep-th/9912001, and references
therein.
\bibitem{HT} S.W. Hawking and N. Turok, Phys. Lett. B {\bf 425}, 25
  (1998); Phys. Lett. B {\bf 432}, 271 (1998).



\end{thebibliography}
\end{document}